\documentstyle[12pt,aaspp4,amssym,flushrt]{article}

\begin{document}

\title{QUASAR-GALAXY AND AGN-GALAXY CROSS-CORRELATIONS}

\author{H\'ector J. Mart\'{\i}nez\altaffilmark{1},
 Manuel E. Merch\'an\altaffilmark{2},  
 Carlos A. Valotto\altaffilmark{2} and
        Diego G. Lambas\altaffilmark{3}}

\affil{Grupo de Investigaciones en Astronom\'{\i}a Te\'orica y Experimental (IATE)}

\affil{Observatorio Astron\'omico de C\'ordoba, Laprida 854, 5000 C\'ordoba,
       Argentina \\
       julian@oac.uncor.edu, manuel@oac.uncor.edu, val@oac.uncor.edu, dgl@oac.uncor.edu}

\altaffiltext{1}{Fa.M.A.F., Universidad Nacional de C\'ordoba, Argentina}
\altaffiltext{2}{On a fellowship from CONICET, Argentina}
\altaffiltext{3}{CONICET, Argentina}

\begin{abstract}

We compute quasar-galaxy and AGN-galaxy cross-correlation functions for samples
taken from the \cite{VCV98} catalog of quasars and active galaxies,
using tracer galaxies taken from the Edinburgh/Durham Southern Catalog.
The sample of active galaxy targets shows positive 
correlation  at projected separations $r_p < 6 h^{-1} ~Mpc$ consistent
with the usual power-law.
On the other hand, we do not find a statistically significant positive 
quasar-galaxy correlation signal except in the range
$3 h^{-1} Mpc < r_p < 6 h^{-1} Mpc$ where we find similar 
AGN-galaxy and quasar-galaxy correlation amplitudes. 
At separations $r_p<3~h^{-1} ~Mpc$ a strong decline of quasar-galaxy 
correlations is observed, suggesting  
a significant local influence of quasars in galaxy formation. 
In an attempt to reproduce the observed cross-correlation between
quasars and galaxies, we have performed CDM cosmological hydrodynamical
simulations and tested the viability of a scenario based on the model
developed by \cite{silkrees98}. 
In this scheme a fraction of the energy released by quasars is
considered to be transferred into
 the baryonic component of the intergalactic medium in the form of winds. 
The results of the simulations suggest that the shape of
the observed quasar-galaxy cross-correlation function could be understood 
in a scenario where a substantial amount of energy is transferred 
to the medium at the redshift of maximum quasar activity.
\end{abstract}


\clearpage

\section{Introduction}

Since the pioneer studies on the cosmological nature of quasar redshifts
(eg. \cite{bsg69}, \cite{gun71}) 
several works have provided useful insights on the environment of quasars
such as the recent works of
\cite{smith95} and \cite{fischer96}.
At low redshifts, $z<0.3$, recent results suggest that quasars are placed
in similar environments to galaxies (\cite{smith95}), or in groups 
of 10-20 galaxies (\cite{fischer96}).
In this respect at high redshifts,
$z>0.6$, the evidence of different environments for radio-loud
and radio-quiet quasars is compelling (\cite{yg87},  
\cite{boy88}, \cite{yee90}). Radio-loud quasars are mainly associated with
rich Abell clusters and radio-quiet quasars reside either
in groups or in the outskirts of clusters.
It should be noted that due to the lack of suitable low
redshift quasar samples the majority
of observational works concern
intermediate redshifts, $0.3<z<0.6$.

On the theoretical side the discussion  of the influence of quasar
activity on the
formation and evolution of nearby galaxies has been present during
the last decades.
The surrounding galaxy environment would be responsible for
triggering and fueling the quasars
as first suggested by \cite{tt72}, where interacting galaxies and mergers
are efficient at transporting gas
into the inner regions of the host galaxy. The evolution of the rate of
galaxy interactions in clusters would 
allow an understanding of the observed clustering of galaxies around
radio-loud quasars (\cite{sp81}).
On the other hand, the large range of influence of a quasar
may have implications for structure formation. According to \cite{rees88}
and \cite{babul91} quasars may ionize the surrounding medium 
up to several megaparsecs away. A different approach is adopted by other
authors (see for instance \cite{voit96}, \cite{nat97}, \cite{silkrees98}),
where a fraction of  the energy released by quasars is transferred
via energetic outflows to the gaseous medium.

In order to shed some light on this subject we have computed
the angular and projected cross-correlation function
of target quasars and AGNs from the \cite{VCV98} 
Catalog, and galaxy tracers from the Edinburgh-Durham Galaxy Catalog 
(\cite{hey89}).  
We have also explored the possible astrophysical effects of quasar activity by
means of
hydrodynamical cosmological numerical simulations which take into
account momentum transfer to the gas at the time
of maximum quasar activity. As an attempt to model these effects we have 
considered here a simple scheme following the model of \cite{silkrees98}.
Quasar-galaxy association in the models are quantified 
by computing the spatial quasar-galaxy cross-correlation function
and we derive the projected cross-correlations 
which may be compared to observations.

\section{Data} \label{data}

Our target sample consists of quasars and AGNs selected from the \cite{VCV98} 
Catalog, hereafter VCV, which comprises angular positions, 
redshifts, and photometric data for 11,358 quasars and 3,334 AGNs and
provides the most recent compilation of quasars and AGNs.

The Edinburgh-Durham Southern Galaxy Catalog, hereafter COSMOS Survey
(\cite{hey89}), was used as a suitable tracer of the galaxy distribution. 
This survey provides angular positions and
photographic magnitudes in the $b_{j}$ band for over two million galaxies. We
have restricted our study to the region $\delta <-5 \arcdeg$ given the
lower quality in the
photographic material in the northern hemisphere.
We adopt a limiting apparent magnitude $b_j=20.5$ as a compromise
between plate quality and depth.
                  
We have estimated the galaxy redshift distribution function for different 
apparent limiting magnitudes as derived by \cite{be93} (see figure 1). 
Taking into account
these results we have adopted a target redshift cutoff $z=0.35$
since the fraction of COSMOS galaxies beyond this redshift is less
than $2 \%$ for limiting magnitude $b_j=20.5$.
We have excluded from our analysis targets with $z < 0.05$. This avoids a few
nearby objects for which fixed projected spatial bins correspond to very large
angular areas dominating projected correlations. 
With the above restrictions the resulting sample 
of targets consists in 103 quasars and 226 AGNs in the region
$\alpha<4\fh 40\fm$ and $\alpha>21\fh$, and $-60\arcdeg<\delta<-5\arcdeg$.
The tracer sample comprises 1,982,847 galaxies brighter than $b_j=20.5$ in the 
same area.

\section{Analysis} \label{methods}

Our analysis consists of the computation of the mean excess with
respect to the average number density of COSMOS galaxies as a function
of angular separation $w(\theta)$), and  projected distance ($W(r_p)$) 
from VCV targets.

\begin{equation}
w(\theta)=\frac{\langle N_{t}(\theta)\rangle}{\langle N_{r}(\theta) \rangle }-1
\label{wcalcang}
\end{equation}

\begin{equation}
W(r_{p})=\frac{\langle N_{t}(r_{p}) \rangle}{\langle N_{r}(r_{p}) \rangle }-1
\label{wcalcpro}
\end{equation}

$\langle N_{t}(\theta) \rangle$, $\langle N_{r}(\theta) \rangle$
,$\langle N_{t}(r_{p}) \rangle$ and $\langle N_{r}(r_{p}) \rangle$
are the mean number of  target-galaxy pairs (VCV-COSMOS) and random-galaxy
pairs (random-COSMOS) at angular separation $\theta$ 
and projected separation $r_p$ respectively.

We have considered cross-correlations for $\theta >10~arcsec $ and
$r_p >0.1~h^{-1}~Mpc$ to exclude the cross-correlation with the host galaxy.
In order to directly compare the results of quasars and AGNs we have
restricted the redshift distribution of AGNs to fit that of quasars. This new 
restricted sample of AGNs has 116 objects and 
was obtained by imposing to the original AGNs sample a radial
gradient through a Monte-Carlo rejection algorithm that provides a similar
redshift distribution to quasars.
Therefore, both angular and projected cross-correlation functions can
be directly compared. 
In figure 1 we show the corresponding redshift distributions of galaxies with 
$b_j<20.5$ (\cite{be93}), quasars, AGNs, and restricted AGNs. As can be 
seen, in this figure a negligible fraction of COSMOS galaxies are expected 
beyond our limiting redshift $z<0.35$.
The different redshift distributions of quasars 
and AGNs and the similar redshift distributions of quasars
and restricted AGNs can also be appreciated.
 
In figures 2 to 5 are shown the resulting angular and projected correlation
for quasars $(w_{Q}(\theta), W_{Q}(r_{p}))$ and the restricted
AGN sample $(w_{A}(\theta), W_{A}(r_{p}))$, both
with COSMOS tracers at limiting magnitude $b_j < 20.5$. In these figures
error bars correspond to uncertainties derived 
from the bootstrap resampling technique
developed by \cite{barrow84} with 20 bootstrap target samples.
Figure 2 shows a lack of strong quasar-galaxy angular correlations
$w_{Q}(\theta)$, 
consistent with the results derived by \cite{smith95}.  
On the other hand the AGN-tracer angular 
cross-correlation function 
$w_{A}(\theta)$ as shown in figure 3 is consistent with a power-law
with significant correlation signal for $\theta < 1,200~arcsec $. 

As can be seen in figure 4 the quasar-tracer projected cross-correlation
function $W_{Q}(r_{p})$ shows a statistically significant 
positive signal for quasar-tracer 
projected separations in the range $3 h^{-1} Mpc < r_p < 6 h^{-1} Mpc$. 
On smaller scales, $r_p<3~h^{-1}~ Mpc$,
there is a marked decrease of the correlation amplitude which makes
$W_{Q}(r_{p}) \simeq 0$.
This behavior contrasts with the smooth power-law 
trend of $W_{A}(r_{p})$ shown in figure 5 for the restricted
sample of AGN targets. 
Thus, a significant difference in the environment of
quasars and active galaxies is implied by our analysis.
While the latter show a 
correlation function similar to that of the general galaxy population,
the projected
quasar-galaxy correlation function, $W_Q(r_p)$, indicates a 
significant departure from a power law at small scales.

By comparing the projected correlation functions $W_Q(r_p)$ and $W_A(r_p)$
it can be seen their similar 
correlation amplitude on scales $3 h^{-1} Mpc<r_p<6h^{-1} Mpc$.
 This result
indicates that quasars and AGNs reside in similar global galaxy density
enhancements and suggesting that the effective influence of 
quasars on galaxy formation is restricted to scales $<3 h^{-1}Mpc$.

\section{Numerical models}

In an attempt to analyze the influence of quasars on environment
we have performed numerical simulations using the Hydra code, developed 
and kindly made available by H. Couchman.
Initial positions and velocities of particles were generated using the Zeldovich
approximation corresponding to the CDM power spectrum.
We have adopted two different values of the density parameter
$\Omega=0.2$ and $\Omega=0.4$ both with baryonic density parameter
$\Omega_B=0.1$.
Simulations  corresponding to $\Omega =0.4$ model were normalized according 
to Cosmic background temperature fluctuations as determined by measurements 
of the COBE satellite, which implies present rms mass fluctuation in a
$8~h^{-1}~Mpc$ sphere radius $\sigma_8=0.75$ (\cite{gorski95}).
We have taken the same $\sigma_8$ value for the $\Omega=0.2$ model
in order to make them of comparable present mass density fluctuations
and to emphasize the dependence of the results on the 
ratio of baryonic-to-total density.
The computational volume corresponds to a periodic cube of side length 
$20 h^{-1} Mpc$ measured at $z=0$.
We have followed the evolution of $N=2 \times 32^3$ particles for
the two CDM models adopted, $\Omega=0.4$ and $\Omega=0.2$ with the same 
value of the Hubble constant 
($H_0= 65~km~s^{-1} Mpc^{-1}$).
The mass of the gas and the dark matter particles are
$2.9\times 10^{9} M_{\sun} $ and $8.59\times 10^{9}M_{\sun}$ respectively
for the $\Omega=0.4$ model and $2.9\times 10^{9} M_{\sun} $ 
and $2.9\times 10^{9}M_{\sun}$ for the $\Omega=0.2$ model. 

We use an analytic fit to the CDM power spectrum of the form

\begin{equation}
p(k) \propto \frac{k}{[1+3.89q+(16.1q)^2+(5.46q)^3+(6.71q)^4]^{1/2}}
\left( \frac{\ln(1+2.34q)}{2.34/q} \right) ^{2}
\end{equation}    
                                 
where $q=(k/\Gamma) h$ Mpc $^{-1}$, $\Gamma=\Omega_{0}~h$ exp
(-$(\Omega_B+
\sqrt{(h/0.5)}\Omega_B/\Omega_0))$ and $\Omega_B=0.0125~h^{-2}$ is the value of 
the baryon density parameter imposed by nucleosynthesis theory 
(\cite{bardeen} and \cite{sugiyama}).
The initial conditions correspond to redshift $z=40$ and
the evolution was followed using a variable time-step resulting typical 
runs of 1000 steps.

We have adopted the start of quasar activity in the models at $z=3$
which coincides approximately with the redshift of observed
maximum abundance of quasars.
At this redshift, we identify local mass density maxima and associate
the quasar with the highest density particle.

In order to analyze the effects of quasar energetics 
on galaxy formation we have 
applied a simple scheme following the model developed by
\cite{silkrees98} in hydrodynamical numerical simulations.
In this plain scheme it is assumed that a fraction ($\sim 10\%$)
of the total energy input by a quasar 
($\sim 10^{62} erg$) is in the form of momentum transfer
to the surrounding gas. This is accomplished by increasing the 
radial component of the velocities of gas particles within a given
radius at $z=3$ where we assume one quasar becomes active 
in each simulation.
According to \cite{silkrees98} (see also \cite{voit96}) this increase
of radial velocities amounts to $\sim 500 
kms^{-1}$ for particles within a sphere of radius 
$R=4.6(E_{62}/\Omega_{0.05})^{1/5}
(1+z)^{-6/5}Mpc$.
Here, $E_{62}$ refers to the total energy in units of $10^{62} erg$ 
emitted by the quasar
in its lifetime ($\sim 10^{7} yr$), and $\Omega_{0.05}$ is the baryonic 
density parameter in units of 0.05. 
The quasar position is chosen to coincide with the highest gas density
particle in the simulations. It should be recalled
that the simulated box is not 
typical regarding the number density of quasars
since to match observed quasar abundance estimates only 1 out of $\sim 100$
such boxes should have a quasar. Our rather small box size adopted
is mainly to improve the numerical resolution of the simulations.

Using a standard friends-of-friends algorithm 
we define gaseous clumps that should be associated to galactic halos  
in the simulations.  
Each given initial condition is evolved with and without quasar activity.
We take advantage of the identical initial conditions
for each pair of simulations for a suitable analysis of 
the environmental effects of quasars. We
compute the spatial quasar-galaxy correlation function in the set of
simulations with quasar energy release and
we compare the quasar-galaxy cross-correlation function with the 
galaxy-galaxy correlation functions of the set of simulations without quasars.
For the latter, the centers of the galaxy-galaxy correlations 
correspond to the maximum gas density particles that defined the quasars
at $z=3$ in the former.

The resulting spatial quasar-galaxy correlation function $\xi_Q(r)$ 
corresponding to 20 pairs of simulations are shown in figures 6 and 7 
($\Omega = 0.2$ and $\Omega = 0.4$ respectively)
which display 
the strong suppression of galaxy formation in the neighborhood of the
quasars ($r<2 h^{-1} Mpc$) consistent with no correlation at these scales,
and a maximum in $\xi_Q(r)$ at $r \simeq 3 h^{-1} Mpc$ since the power-law
behavior at larger scales is not affected by the model.
For comparison, in these figures is also shown the spatial galaxy-galaxy
correlation function with center galaxies defined in the previous paragraph.
In both figures, errorbars correspond to Poisson estimates, $\delta \xi_Q =
\sqrt {N}/ N_u $ where $N$ is the number of
galaxies in each bin and $N_u$ is the expected number in a uniform distribution.

Comparing the spatial correlation functions of the
models a stronger galaxy suppression at small
scales is seen in the lower density ($\Omega = 0.2$) model. This is due to the
higher relative fraction of baryons which are affected by the
momentum transfer in this case. In the $\Omega =0.4$ model 
the dark matter is dominant by a factor 3 allowing the retention of more gas  
in the quasar neighborhood than in the low density models. 

In figures 8 and 9, we show the projected quasar-galaxy
correlation function of the models $W(r_p)$. In the small angle
approximation $W_Q(r_p)$ is related to the spatial correlation function
$\xi_Q(r)$ by:

\begin{equation}
W_{Q}(r_p) =  B \int^{\infty}_{-\infty}\xi_Q((r_p^{2}+\pi^{2})^{1/2})d\pi
\end{equation}

with $B = \sum_i p(y_i)/(\sum_i(1/y_i^2)\int_0^{\infty} p(x) x^2 dx)$
is a normalization constant which includes the probability $p(x)$
that a tracer galaxy is at distance $x$, and $y_i$ is the distance to the 
$i^{th}$ quasar.
We have used a Schechter function fit (\cite{schter76}) to the
luminosity function of COSMOS tracer galaxies, with parameters
$M^*=-19.7$, $\alpha=-1.0$, a flat cosmology, and a linear $K$-correction
term $K=3 z$.

The model projected quasar-galaxy correlation function
 on scales $r_p < 3 ~h^{-1}~ Mpc$
has a similar behavior than the observed cross-correlation. 
By inversion of equation 4
using the observed $W_Q(r_p)$  we derive negative values of
the spatial cross-correlation function $\xi_Q(r)$  
on these scales, which
stress the relevance of this observational effect.
 
The simplicity of our adopted scheme of energy release of quasars, 
the small size of the simulated box and the restricted dynamical range
impose further limitations on the models. Considering these shortcomings,
we conclude that the results of the models analyzed reasonably reproduce the 
observations.

\section{Conclusions} 

The quasar-galaxy projected
cross-correlation function $W_{Q}(r_p)$ shows a 
maximum at $\simeq 3 h^{-1} Mpc$
and is consistent with no correlation for $r_p < 3 h^{-1} Mpc$.
The inversion of such a projected correlation function requires a negative
spatial correlation on scales $\leq 2 h^{-1} Mpc$ indicating
strong effects of quasars in their local environment.
The corresponding angular cross-correlations are
roughly consistent with no signal detection.
In contrast to the behavior of quasar-tracer correlations,
we find that both angular and projected AGN-galaxy cross-correlation
functions are consistent with the usual power-law shape.
The agreement of $W_Q(r_p)$ and $W_A(r_p)$ at scales
$3~h^{-1}~Mpc<r_p<6~h^{-1}~Mpc$, suggests that both quasars
and AGNs reside in similar global galaxy enhancements.

The analysis of cosmological hydrodynamical simulations
allows us to study the effects
of energy input by the quasars to the surrounding gaseous medium.
The model developed by \cite{silkrees98} considers a typical 
energy input of $\sim 10^{61}~erg $ to the  
gas which develops a shell of present radius $\sim 4 Mpc$ originally
at a speed of $\sim 500~ km~s^{-1}$. 
The implementation of this scheme in our numerical simulations
shows that a relative lack of galaxies in the neighborhood of
quasars may arise naturally in these models providing support to the
viability of this scenario.

\acknowledgments
This work was partially supported by the Consejo de 
Investigaciones Cient\'{\i}ficas y
T\'ecnicas de la Rep\'ublica Argentina, CONICET, the Consejo de
 Investigaciones
Cient\'{\i}ficas y Tecnol\'ogicas de la Provincia de C\'ordoba, CONICOR and 
Fundaci\'on Antorchas.

\clearpage

\figcaption{ Redshift distributions. The thick solid line corresponds to 
tracer COSMOS galaxies with $b_j < 20.5$ according to \cite{be93}, 
the thin solid line corresponds to quasars, dashed lines to AGNs, and
dotted lines to restricted AGNs.}

\figcaption{Quasar-tracer angular cross-correlation function $w_Q(\theta)$. 
Tracer galaxies correspond to COSMOS survey with $b_j < 20.5$.}

\figcaption{AGN-tracer angular cross-correlation function $w_A(\theta)$. 
Tracer galaxies correspond to COSMOS survey with $b_j < 20.5$.}

\figcaption{Quasar-tracer projected cross-correlation function $W_Q(r_p)$. 
Tracer galaxies correspond to COSMOS survey with $b_j < 20.5$.}

\figcaption{AGN-tracer projected cross-correlation function $W_A(r_p)$. 
Tracer galaxies correspond to COSMOS survey with $b_j < 20.5$.}

\figcaption{Circles: spatial quasar-galaxy cross-correlation function
 $\xi_{Q}(r)$ in the $\Omega=0.2$ models.
 Squares correspond to the galaxy-galaxy correlation function in models
with no quasar activity as described in the text.}

\figcaption{Spatial quasar-galaxy cross-correlation function $\xi_{Q}(r)$
in the $\Omega=0.4$ numerical simulations. All symbols are as in figure 6.}

\figcaption{Projected quasar-galaxy cross-correlation function $W_{Q}(r_p)$
for the $\Omega=0.2$ numerical simulations. Solid and dashed lines correspond 
to models with and without quasar activity respectively.}

\figcaption{Projected quasar-galaxy cross-correlation function $W_{Q}(r_p)$
for the $\Omega=0.4$ numerical simulations. Lines represent the
same as in figure 8.}

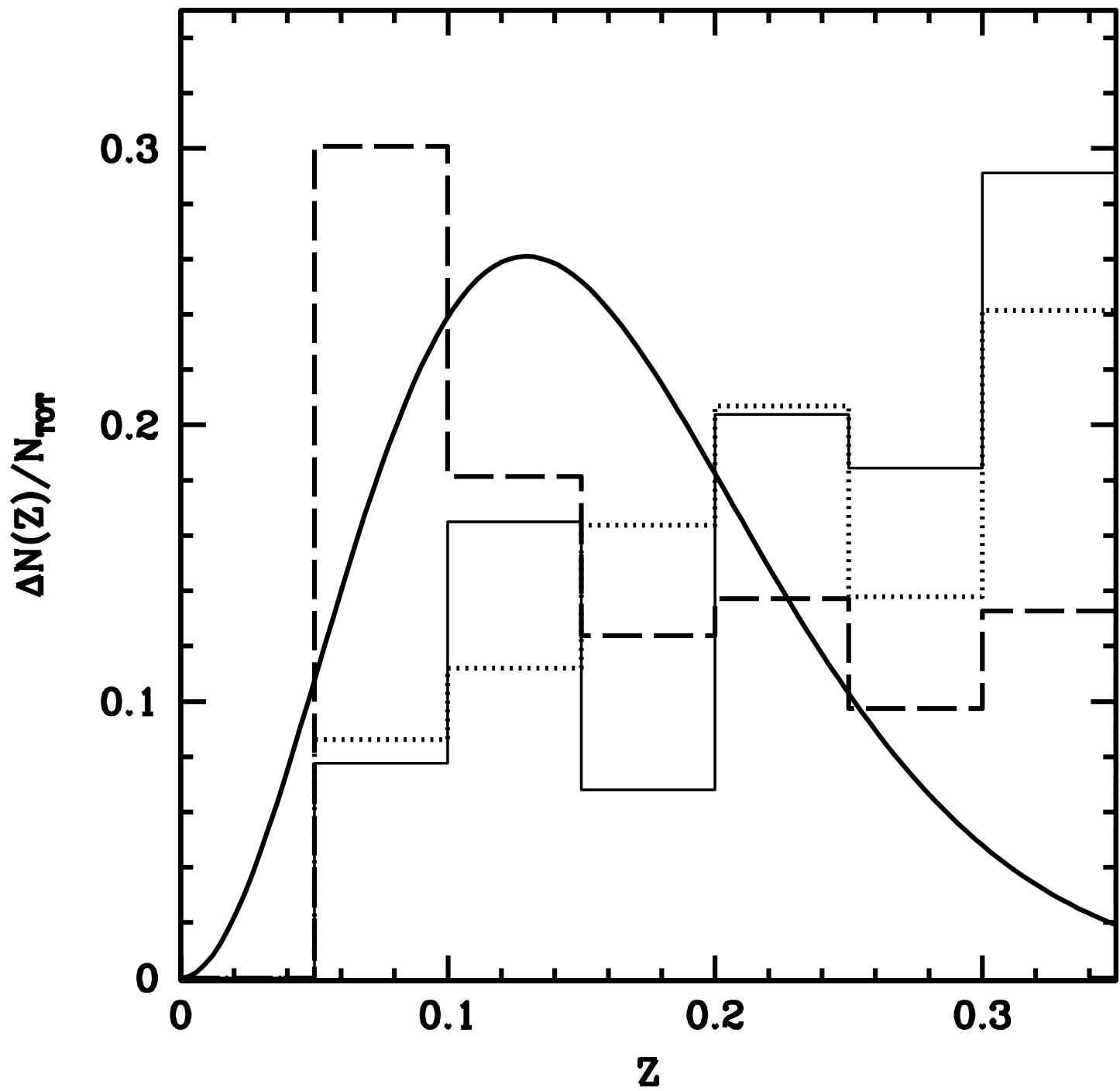
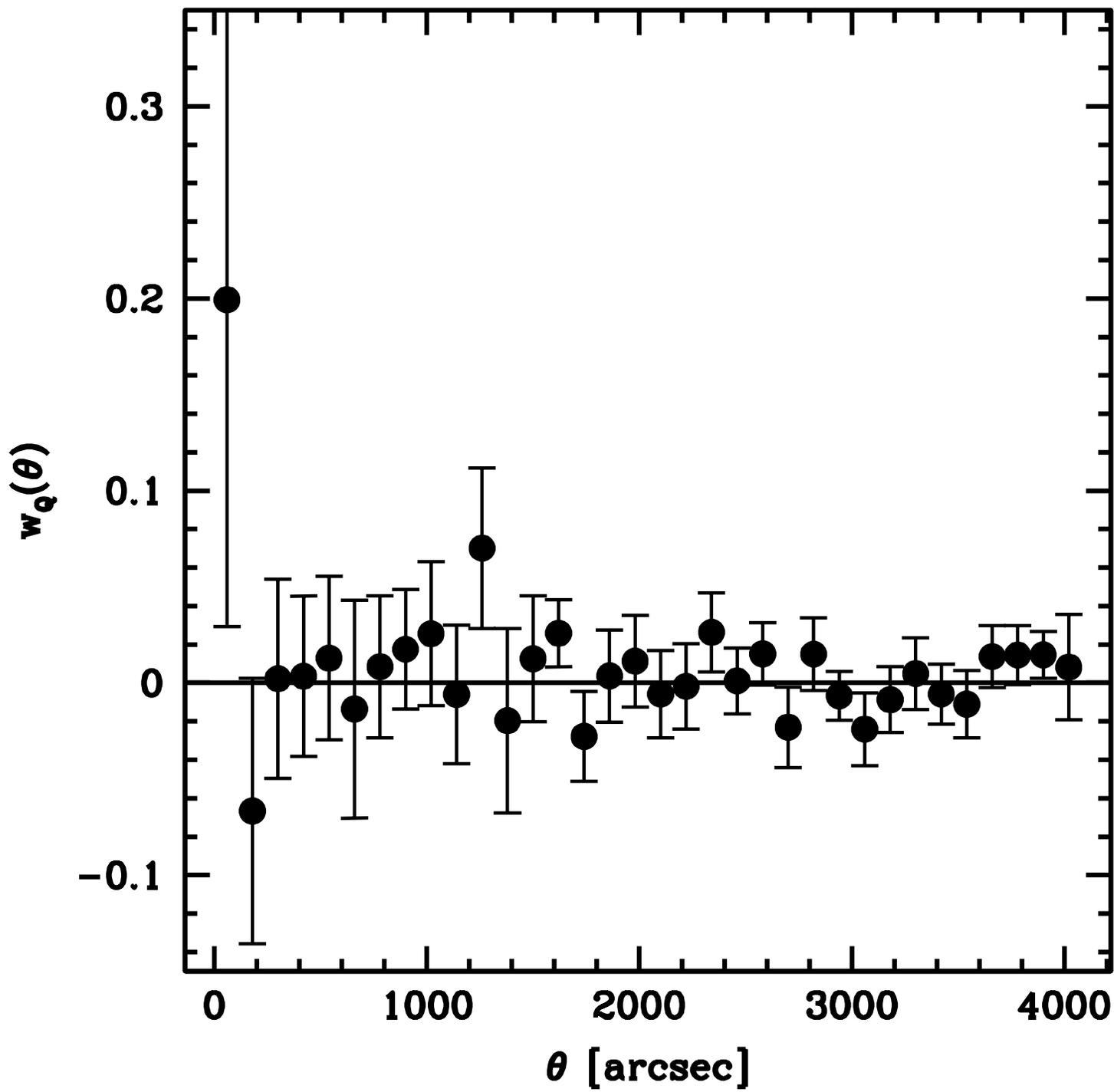
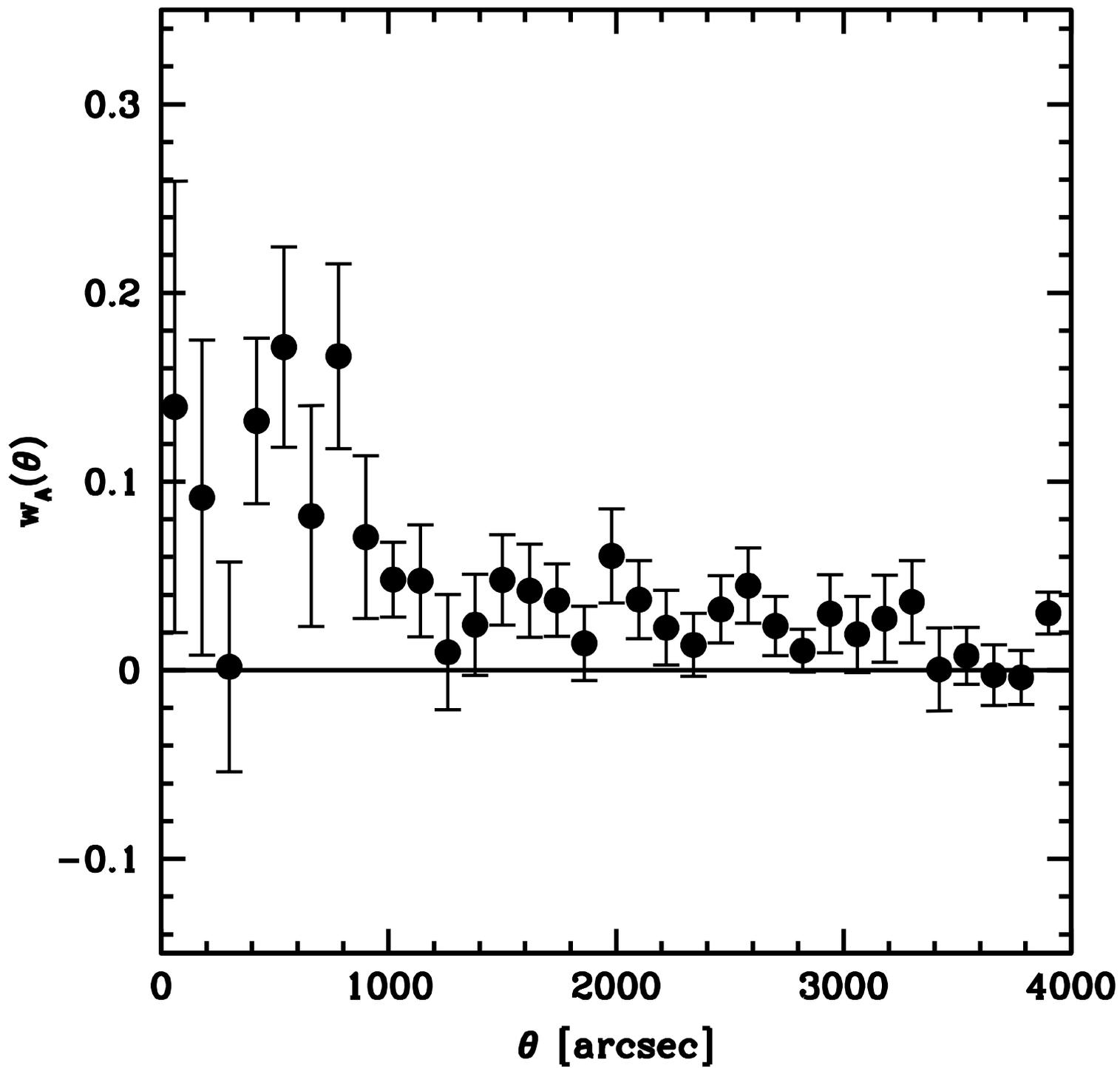
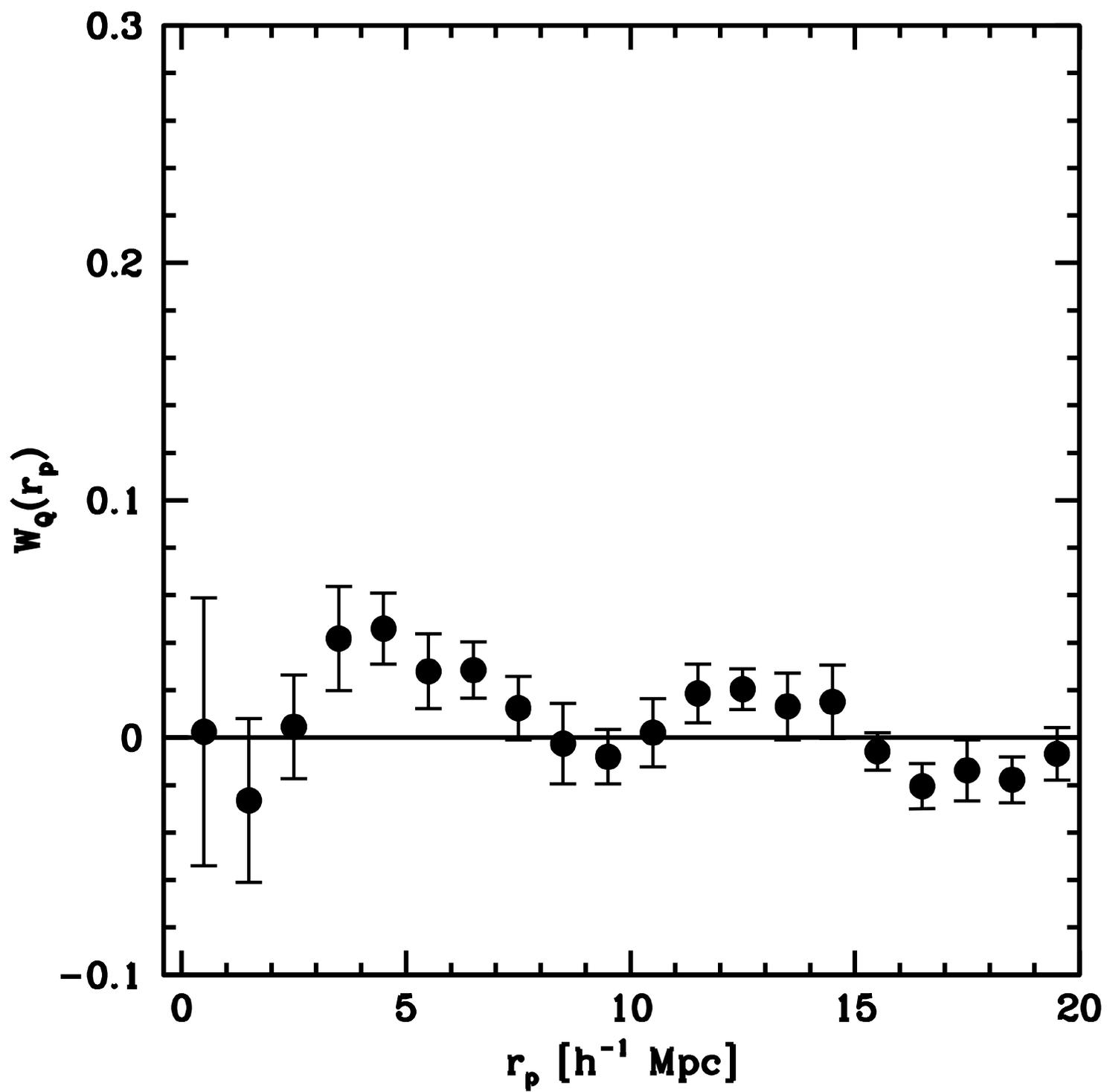
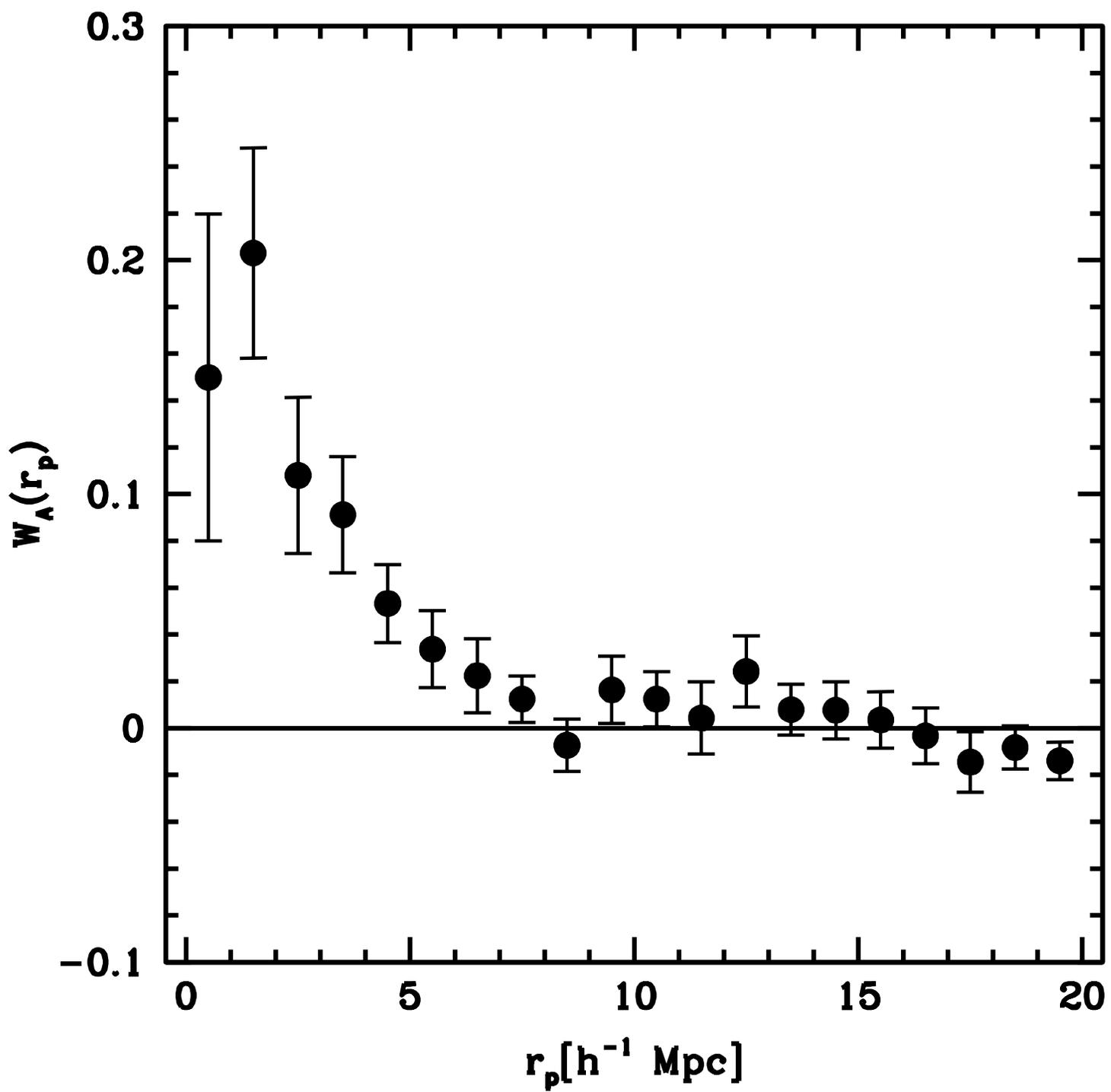
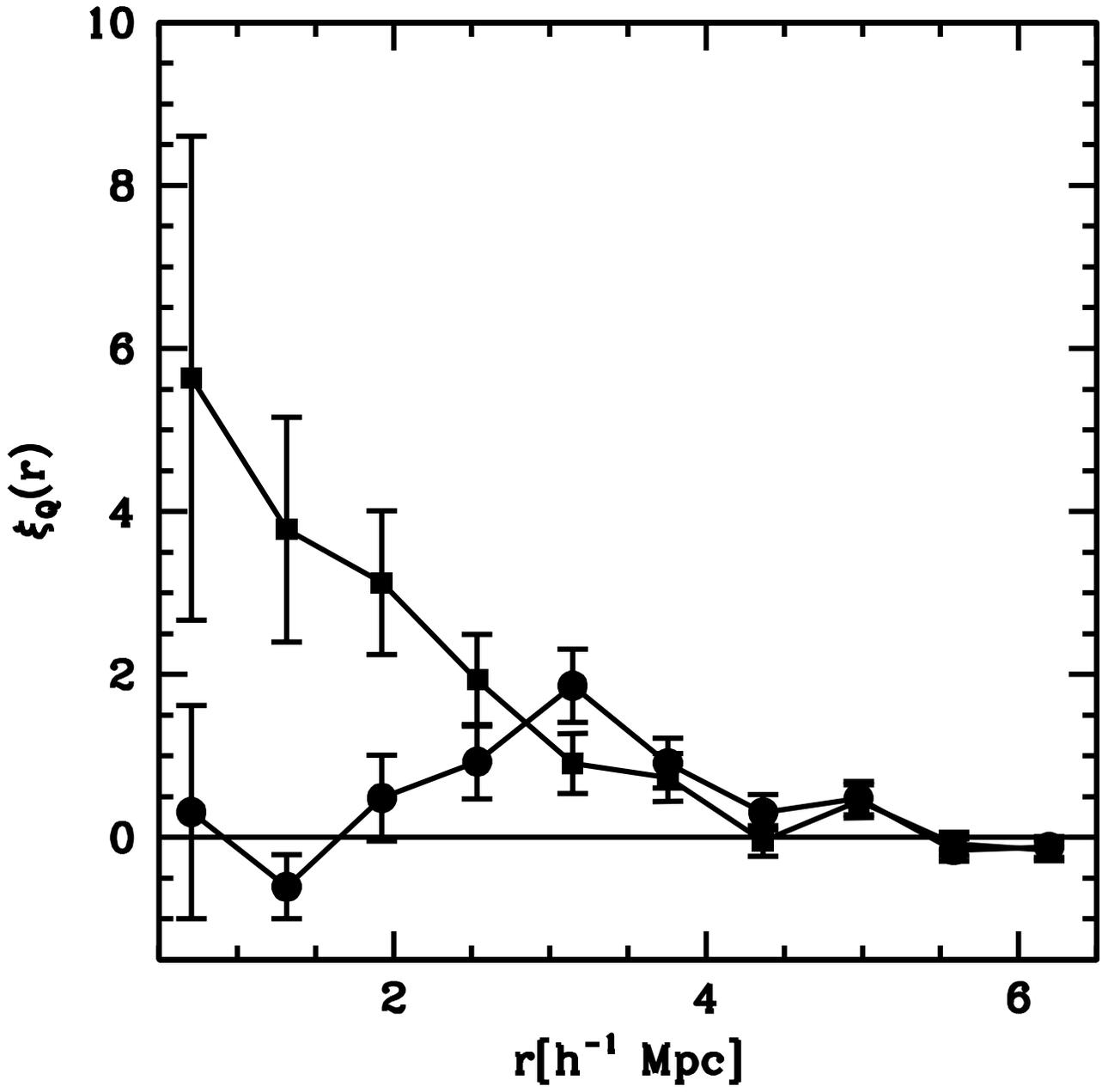
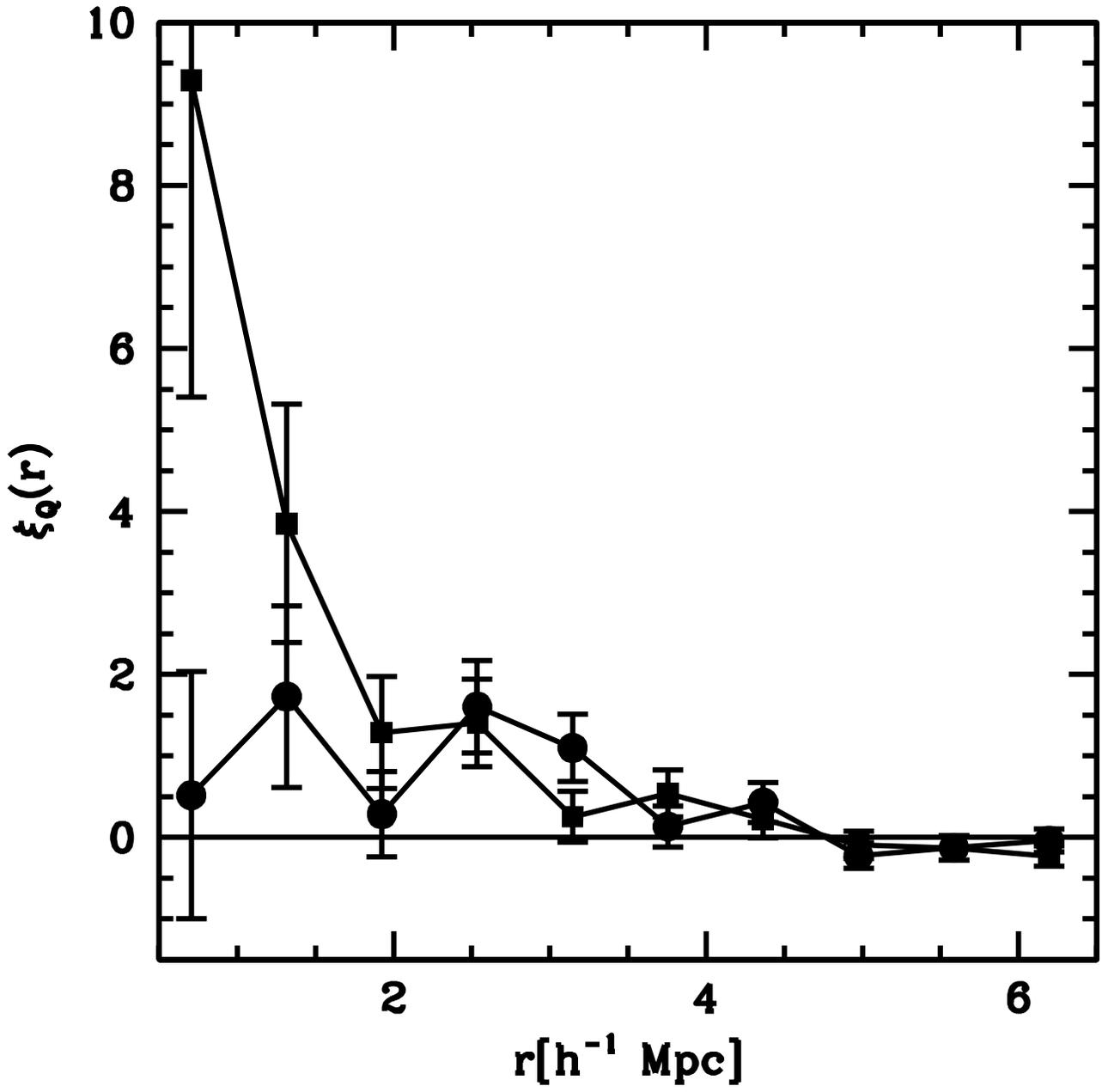
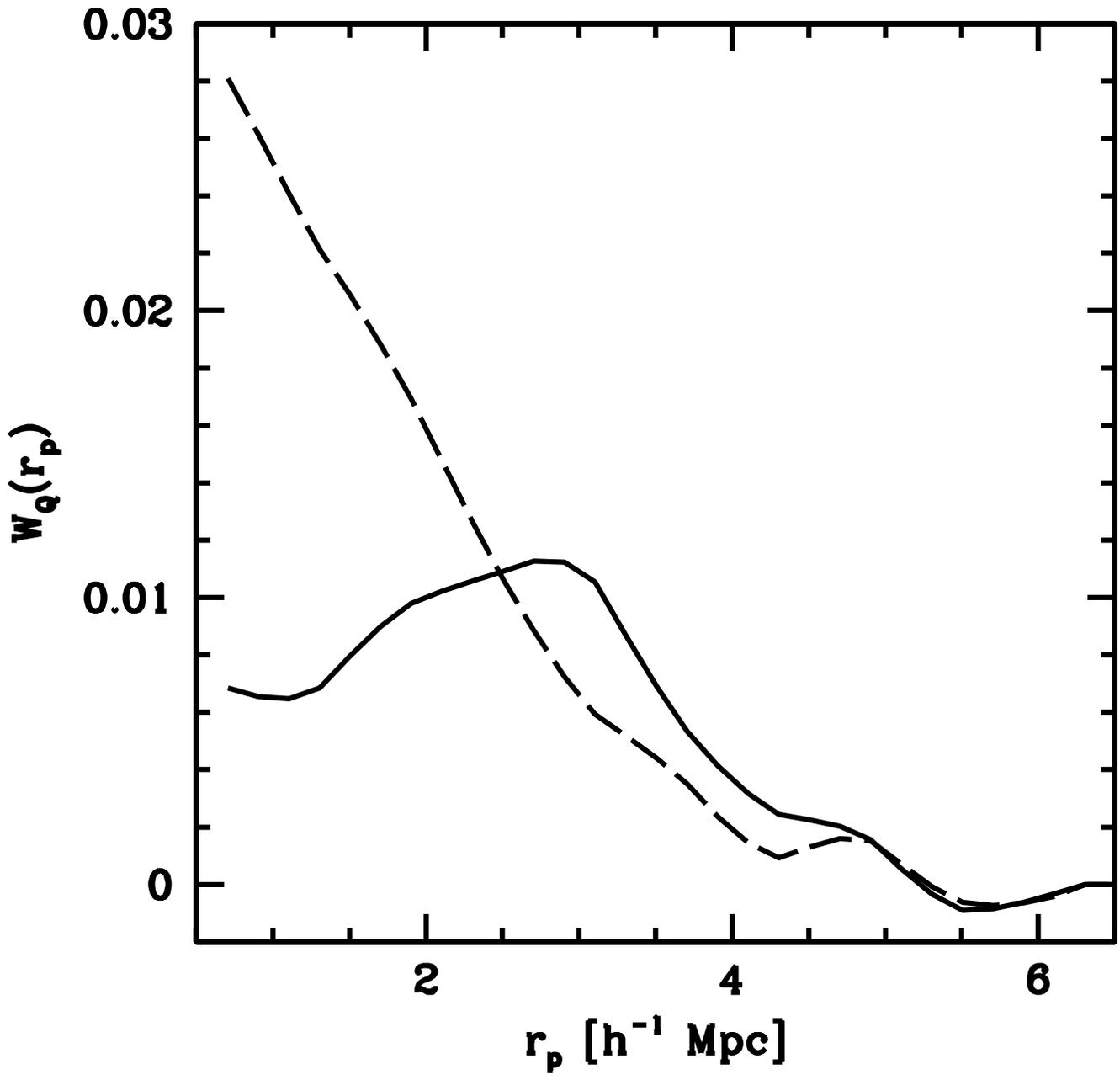
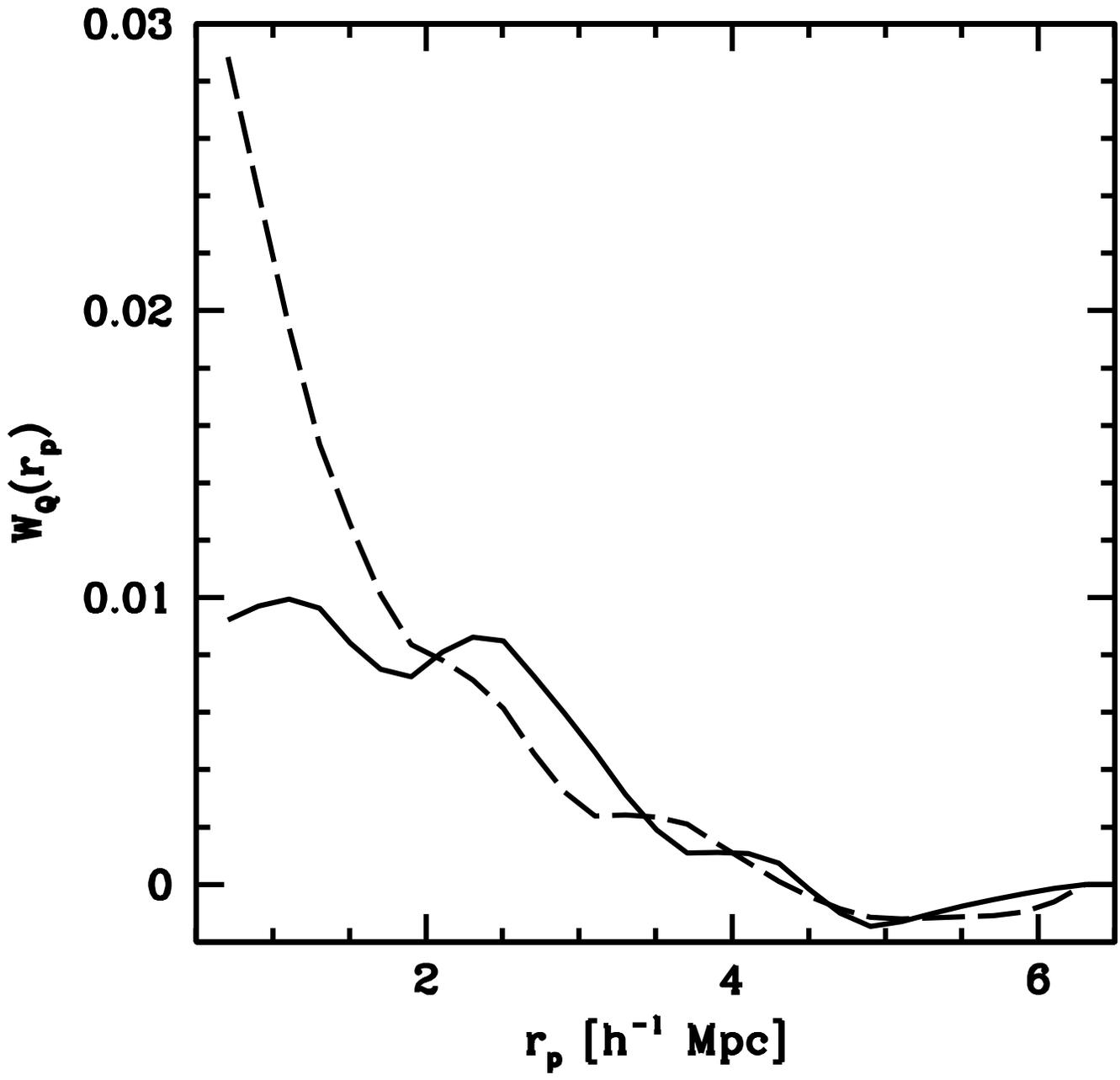
\end{document}